\documentclass[12pt]{article}

\usepackage{amsmath}
\usepackage{graphicx}

\newcommand{\ket}[1]{|\, #1 \rangle}
\newcommand{\bra}[1]{\langle #1 \,|}
\newcommand{\udt}[3]{#1^{#2}_{\phantom{#2}#3}}
\newcommand{\dut}[3]{#1_{#2}^{\phantom{#2}#3}}

\begin{document}

\begin{center}
{\Large\bf Spin Decoherence in a Gravitational Field}
\vskip .6 cm
Hiroaki Terashima$^{1,2}$ and Masahito Ueda$^{1,2}$
\vskip .4 cm
{\it $^1$Department of Physics, Tokyo Institute of Technology,\\
Tokyo 152-8551, Japan} \\
{\it $^2$CREST, Japan Science and Technology Corporation (JST),\\
Saitama 332-0012, Japan}
\vskip .6 cm
\end{center}

\begin{abstract}
We discuss a mechanism of spin decoherence
in gravitation within the framework of general relativity.
The spin state of a particle moving in a gravitational field
is shown to decohere due to the curvature of spacetime.
As an example, we analyze a particle going around
a static spherically-symmetric object.
\end{abstract}

\begin{flushleft}
{\footnotesize
}
\end{flushleft}

\section{Introduction}
The spin of a particle is an interesting degree of freedom
in quantum theory.
Recently, Peres, Scudo, and Terno~\cite{PeScTe02}
have shown that in special relativity
the spin entropy
(i.e., the von Neumann entropy of a spin state)
of a particle is not invariant
under Lorentz transformations
unless the particle is in a momentum eigenstate.
Namely, even if the spin state is
pure in one frame of reference,
it may become mixed in another frame of reference.
The origin of this spin decoherence is that
the Lorentz transformation entangles
the spin and momentum
via the Wigner rotation~\cite{Wigner39}.
The entanglement then produces spin entropy
by a partial trace over the momentum.

In this paper, we study the spin state of a particle
moving in a gravitational field to show its decoherence
by the effects of general relativity~\cite{TerUed03c}.
Our result implies that
even if the spin state is pure at one spacetime point,
it may become mixed at another spacetime point.
This spin decoherence is derived from
the curvature of spacetime caused by
the gravitational field.
Such a spacetime curvature entails
a local description of spin by local Lorentz frame
due to a breakdown of the global rotational symmetry.
The motion of a particle is thus accompanied by
the change of frame, which can increase the spin entropy
analogous to the case of special relativity.
As an example, we consider a particle in a circular
orbit around a static spherically-symmetric object
using the Schwarzschild spacetime.

\section{Formulation}
Consider a wave packet of a spin-$1/2$ particle
with mass $m$ in a gravitational field.
The gravitational field is described by
a curved spacetime with metric
in general relativity.
Nevertheless, despite the spacetime curvature,
we can locally describe this wave packet
as if it were in a flat spacetime,
since a curved spacetime locally looks like flat.
More precisely, for any spacetime point
we can find a coordinate system in which
the metric becomes the Minkowski one.
The coordinate transformation
from a general coordinate system $\{x^\mu\}$
to this local Lorentz frame $\{x^a\}$
can be carried out using a vierbein (or a tetrad)
$\dut{e}{a}{\mu}(x)$ and
its inverse $\udt{e}{a}{\mu}(x)$
defined by~\cite{BirDav82}
\begin{align}
\dut{e}{a}{\mu}(x)\,\dut{e}{b}{\nu}(x)
 \,g_{\mu\nu}(x) &= \eta_{ab}, \notag \\
\udt{e}{a}{\mu}(x)\,\dut{e}{b}{\mu}(x)
       &=\udt{\delta}{a}{b},
\label{eq:vierbein}
\end{align}
where $g_{\mu\nu}(x)$ is the metric
in the general coordinate system
and $\eta_{ab}\equiv\mathrm{diag}(-1,1,1,1)$
is the Minkowski metric with $a,b=0,1,2,3$.
The vierbein then transforms a tensor
in the general coordinate system
into that in the local Lorentz frame, and vice versa.
For example, momentum $p^\mu(x)$
in the general coordinate system can be transformed
into that in the local Lorentz frame
via the relation $p^a(x)=\udt{e}{a}{\mu}(x)\, p^\mu(x)$.

Therefore,
we describe the wave packet
as in the case of special relativity~\cite{PeScTe02}
using a local Lorentz frame
at the spacetime point $x^\mu$ where
the centroid of the wave packet is located.
Since a momentum eigenstate
of the particle is labeled by four-momentum
$p^a=(\sqrt{|\vec{p}|^2+m^2c^2},\vec{p})$ and
by the $z$-component $\sigma$ ($=\uparrow$, $\downarrow$)
of spin~\cite{Weinbe95} as $\ket{p^a,\sigma}$,
the wave packet can be expressed by
a linear combination
\begin{equation}
 \ket{\psi}=\sum_\sigma \int d^3\vec{p}\;N(p^a)
 \,C(p^a,\sigma) \,\ket{p^a,\sigma},
\end{equation}
where
\begin{equation}
d^3\vec{p}\;N(p^a)\equiv
d^3\vec{p}\;\frac{mc}{\sqrt{|\vec{p}|^2+m^2c^2}}
\end{equation}
is a Lorentz-invariant volume element.
From the normalization condition
\begin{equation}
\langle p'^a,\sigma'\,|\,p^a,\sigma \rangle=\frac{1}{N(p^a)}\,
\delta^3(\vec{p'}-\vec{p})\,\delta_{\sigma'\sigma},
\end{equation}
the coefficient $C(p^a,\sigma)$ must satisfy
\begin{equation}
  \sum_\sigma \int d^3\vec{p}\;N(p^a)
   |C(p^a,\sigma)|^2=1.
\end{equation}

To obtain the spin state of this wave packet,
we take the trace of the density matrix
$\rho=|\psi\rangle\langle\psi|$ over the momentum,
\begin{equation}
  \rho_\mathrm{r}(\sigma';\sigma)
  = \int d^3\vec{p}\;N(p^a)\,
      \bra{p^a,\sigma'}\rho\ket{p^a,\sigma}.
\end{equation}
The spin entropy is then given by
the von Neumann entropy of
this reduced density matrix:
\begin{equation}
  S=-\mathrm{Tr}\left[\rho_\mathrm{r}(\sigma';\sigma)
     \log_2\rho_\mathrm{r}(\sigma';\sigma)\right].
\end{equation}
Moreover, the spin operator for the wave packet is defined by
\begin{equation}
\hat{\vec{S}}=\frac{1}{2}\sum_{\alpha,\beta}\vec{\sigma}_{\alpha\beta}
  \int d^3\vec{p}\;N(p^a) \,\ket{p^a,\alpha} \bra{p^a,\beta},
\end{equation}
with $\vec{\sigma}$ being the Pauli matrices.
Note that this spin is not Dirac spin
but Wigner one~\cite{Terno02}.
As is well known, Dirac spin, which corresponds to
the index of $4$-component Dirac spinor,
is not a conserved quantity in a relativistic regime
and thus is not suitable degree of freedom
for labelling one-particle states.
In contrast,
Wigner spin is a conserved quantity
suitable for labelling one-particle states,
because it is defined using the particle's rest frame.

\section{Decoherence}
Suppose that the centroid of the wave packet
is moving with four-velocity $u^\mu(x)$
normalized as $u^\mu(x)u_\mu(x)=-c^2$;
this motion is not necessarily geodesic
in the presence of an external force.
After an infinitesimal proper time $d\tau$,
the centroid moves to
a new point $x'^\mu=x^\mu+u^\mu(x)d\tau$
and then the wave packet is described
by the local Lorentz frame at the new point.
This change in the local Lorentz frame
is represented by a Lorentz transformation
$\udt{\tilde{\Lambda}}{a}{b}(x)=
\udt{\delta}{a}{b}+\udt{\chi}{a}{b}(x)d\tau$, where
\begin{equation}
\udt{\chi}{a}{b}(x)
=u^\mu(x) \left[\,\dut{e}{b}{\nu}(x)\nabla_\mu
    \udt{e}{a}{\nu}(x)\,\right].
\end{equation}
In addition to this change,
the acceleration by an external force is also
interpreted as a Lorentz transformation.
Thus, the motion of the wave packet is
equivalent to a Lorentz transformation
$\udt{\Lambda}{a}{b}(x)=\udt{\delta}{a}{b}
+\udt{\lambda}{a}{b}(x)d\tau$,
where~\cite{TerUed03b}
\begin{align}
& \udt{\lambda}{a}{b}(x)=\udt{\chi}{a}{b}(x) \notag \\
& \quad-\frac{1}{mc^2}\left[\,a^a(x)\,q_b(x)
 -q^a(x)\,a_b(x)\,\right]
\label{eq:ill}
\end{align}
using the momentum and acceleration of the centroid
in the local Lorentz frame
\begin{align}
 q^a(x)&=\udt{e}{a}{\mu}(x)\left[\,m u^\mu(x)\,\right],
  \label{eq:momcen} \\
 a^a(x)&=\udt{e}{a}{\mu}(x)\,\left[\,u^\nu(x)
 \nabla_\nu u^\mu(x)\,\right].
\end{align}
Note that even if the wave packet moves
as straight as possible along a geodesic curve,
this Lorentz transformation may be nonzero
in general relativity because of the first term.

Since spin entropy is not invariant
under a Lorentz transformation~\cite{PeScTe02},
neither is it invariant during
the motion of the wave packet.
Note that a Lorentz transformation
rotates the spin of a particle through
an angle that depends on the particle's momentum;
this rotation is known as Wigner rotation.
The momentum eigenstate $\ket{p^a,\sigma}$ thus
transforms under the Lorentz transformation
(\ref{eq:ill}) as~\cite{Weinbe95}
\begin{align}
& U(\Lambda(x))\, \ket{p^a,\sigma} \notag \\
& \qquad=\sum_{\sigma'} D_{\sigma'\sigma}
 (W(x))\,\ket{\Lambda(x) p^a,\sigma'},
\end{align}
where $D_{\sigma'\sigma}(W(x))$ is
the $2\times 2$ unitary matrix that represents
a Wigner rotation given by~\cite{TerUed03b}
\begin{align}
 \udt{W}{i}{k}(x) &=\udt{\delta}{i}{k}
  +\udt{\lambda}{i}{k}(x)\,d\tau \notag \\
 & +\frac{\udt{\lambda}{i}{0}(x)\,p_k-
   \lambda_{k0}(x)\,p^i}{p^0+mc}\,d\tau
 \label{eq:deflw}
\end{align}
with $i,k=1,2,3$.
Taking the trace of the density matrix
\[ \rho'=U(\Lambda(x))|\psi\rangle
\langle\psi|U(\Lambda(x))^\dagger \]
over the momentum,
we obtain the spin state
$\rho'_\mathrm{r}(\sigma';\sigma)$
and the spin entropy $S'$ of the wave packet
in the local Lorentz frame at the new point $x'^\mu$.
However, the spin has been entangled with the momentum
by the Lorentz transformation (\ref{eq:ill}),
since the Wigner rotation (\ref{eq:deflw}) of spin
depends on the momentum.
Due to this entanglement,
the new entropy $S'$ is not,
in general, equal to the original one $S$.
This implies that the spin state
may decohere during the motion of the wave packet
by the effects of general relativity.

\section{Example}
As an example in general relativity,
we consider the Schwarzschild spacetime,
which is the unique static spherically-symmetric
solution of Einstein's equation in vacuum.
In the spherical coordinate system
$(t,r,\theta,\phi)$, the metric is given by
\begin{align}
 g_{\mu\nu}(x)dx^\mu dx^\nu
  &=-f(r)c^2dt^2+\frac{1}{f(r)}dr^2 \notag \\
  & \;+r^2(d\theta^2+\sin^2\theta d\phi^2),
\end{align}
where $f(r)=1-(r_s/r)$
with the Schwarzschild radius $r_s$.
In the Schwarzschild spacetime,
we introduce an observer at each point
who is static
with respect to the time $t$
using a static local Lorentz frame.
The vierbein (\ref{eq:vierbein}) is then
\begin{align}
 & \dut{e}{0}{t}(x)=\frac{1}{c\sqrt{f(r)}},\quad
\dut{e}{1}{r}(x)=\sqrt{f(r)} , \notag \\
 & \dut{e}{2}{\theta}(x)=\frac{1}{r}, \quad
\dut{e}{3}{\phi}(x)=\frac{1}{r\sin\theta}.
\end{align}

\begin{figure}
\begin{center}
\includegraphics[scale=0.8]{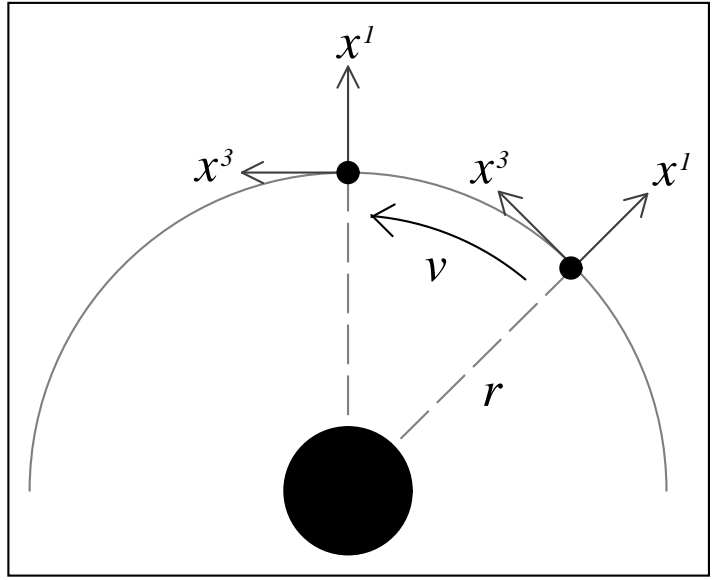}
\end{center}
\caption{\label{fig1}A wave packet (small circle) going around
a static spherically-symmetric object (large circle).
The $1$- and $3$-axes of the local Lorentz frame
are illustrated at the initial and final points.}
\end{figure}

Suppose that the centroid of the wave packet
is moving along a circular trajectory of
radius $r$ ($>r_s$) with a constant velocity
$rd\phi/dt\equiv v\sqrt{f(r)}$ on the equatorial plane
$\theta=\pi/2$ (see Fig.~\ref{fig1}).
The four-velocity of the centroid is then given by
\begin{equation}
 u^t(x)=\frac{\cosh\xi}{\sqrt{f(r)}}, \quad
 u^\phi(x)=\frac{c\sinh\xi}{r},
\end{equation}
where $\xi$ is defined by $\tanh\xi=v/c$.
We assume that at the initial point
the wave packet has
the definite values $\sigma=\uparrow$ and $p^1=p^2=0$
but is Gaussian in $p^3$ with width $w$:
\begin{align}
 &|C(p^a,\sigma)|^2 =\frac{1}{N(p^a)}\,
 \delta(p^1)\,\delta(p^2)\,
 \delta_{\sigma,\uparrow}  \notag \\
 & \qquad\;\times \frac{1}{\sqrt{\pi}w}
 \exp\left[-\frac{(p^3-q^3(x))^2}{w^2}\right],
\end{align}
where $q^3(x)=mc\sinh\xi$ is the momentum of the centroid
(\ref{eq:momcen}) along the $3$-direction.

\begin{figure}
\begin{center}
\includegraphics[scale=0.8]{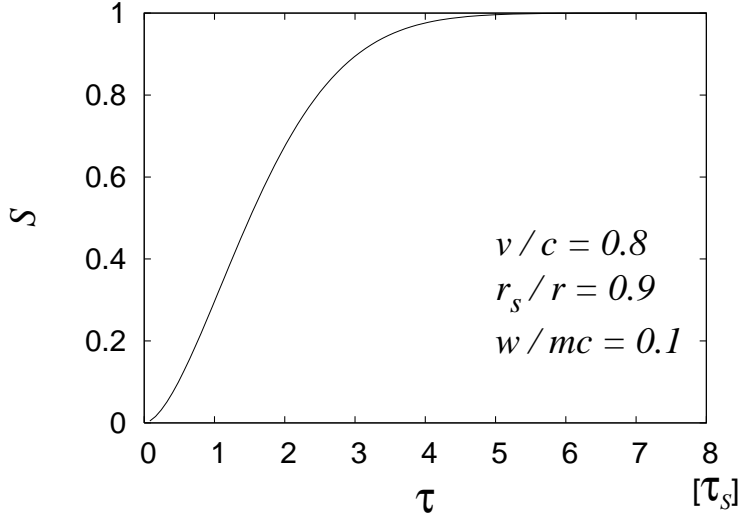}
\end{center}
\caption{\label{fig2}The spin entropy $S$ at
$v/c=0.8$, $r/r_s=0.9$, and $w/mc=0.1$
as a function of the proper time $\tau$
normalized by $\tau_s\equiv mr_s/w$.}
\end{figure}

\begin{figure}
\begin{center}
\includegraphics[scale=0.8]{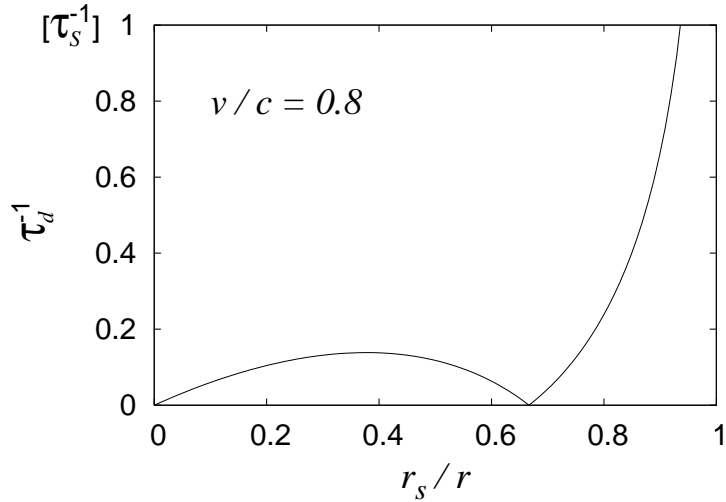}
\end{center}
\caption{\label{fig3}The inverse of the characteristic
decoherence time $\tau_d^{-1}$,
normalized by $\tau_s^{-1}$,
as a function of $r_s/r$ at $v/c=0.8$.}
\end{figure}

Clearly, the spin entropy of this wave packet is
zero at the initial point,
since the spin is separable from the momentum.
However, after a proper time $\tau$ of the particle,
spin entropy is generated
by the gravity and acceleration, i.e.,
by the first and second terms in Eq.~(\ref{eq:ill}).
Figure~\ref{fig2} shows the generated spin entropy $S$
as a function of the proper time $\tau$
in the case of $v/c=0.8$, $r/r_s=0.9$, and $w/mc=0.1$.
The spin state of the wave packet
decoheres to a mixed state
and becomes maximally mixed ($S\to 1$)
in the limit of $\tau=\infty$.
The characteristic decoherence time
is given by the inverse of
\begin{equation}
 \tau_d^{-1}\equiv\frac{w\left(\cosh\xi-1\right)}{mr}
   \left| 1-\frac{r_s}{2rf(r)}\right|
   \sqrt{f(r)}.
\end{equation}
Figure~\ref{fig3} shows this value
$\tau_d^{-1}$
as a function of $r_s/r$ when $v/c=0.8$.
No decoherence occurs ($\tau_d\to\infty$)
at the spatial infinity $r\to\infty$,
whereas extremely rapid decoherence
occurs ($\tau_d\to0$)
near the Schwarzschild radius $r\to r_s$.
The spin state does not decohere
also at $r=3r_s/2$,
because the first term in Eq.~(\ref{eq:ill})
is canceled by the second term.

Of course, this spin decoherence is very slow
in the gravitational field of the earth
$r_s\sim 1 \,\textrm{cm}$.
For example, when a wave packet is at rest in
the International Space Station
going around the earth,
$v \sim 7.7 \,\textrm{km/s}$ and
$r \sim 6800 \,\textrm{km}$,
the characteristic decoherence time
is $\tau_d\sim 2.2\times mc/w$ years.

\section{Conclusion}
We have shown that spin entropy
is generated when a particle
moves in a gravitational field.
The spin state evolves into a mixed state
even if the particle moves
as straight as possible along a geodesic curve.
This decoherence is due to
the spacetime curvature by gravity.

\section*{Acknowledgments}
This research was supported by a Grant-in-Aid
for Scientific Research (Grant No.~15340129) by
the Ministry of Education, Culture, Sports,
Science and Technology of Japan.


\begin{thebibliography}{10}

\bibitem{PeScTe02}
A. Peres, P.~F. Scudo, and D.~R. Terno, Phys. Rev. Lett. {\bf 88},  230402
  (2002).

\bibitem{Wigner39}
E.~P. Wigner, Ann. Math. {\bf 40},  149  (1939).

\bibitem{TerUed03c}
H. Terashima and M. Ueda, J. Phys. A: Math. Gen. {\bf 38},  2029  (2005).

\bibitem{BirDav82}
N.~D. Birrell and P.~C.~W. Davies, {\em Quantum Fields in Curved Space}
  (Cambridge University Press, Cambridge, 1982).

\bibitem{Weinbe95}
S. Weinberg, {\em The Quantum Theory of Fields} (Cambridge University Press,
  Cambridge, 1995).

\bibitem{Terno02}
D.~R. Terno, Phys. Rev. A {\bf 67},  014102  (2003).

\bibitem{TerUed03b}
H. Terashima and M. Ueda, Phys. Rev. A {\bf 69},  032113  (2004).

\end{thebibliography}
\end{document}